\documentstyle[12pt]{article}
\def\beq{\begin{equation}}
\def\eeq{\end{equation}}

\begin{document}
\begin{titlepage}
\begin{center}
{\Large \bf Properties of perturbative multi-particle amplitudes in
$\phi^k$ and $O(N)$ theories\\} \vspace{0.2in}
{\bf B.H. Smith  \\ }
School of Physics and Astronomy \\
University of Minnesota \\
Minneapolis, MN 55455 \\

\vspace{0.2in}
{\bf   Abstract  \\ }
\end{center}
Threshold amplitudes are considered for multi-particle production in
$\phi^k$ and $O(N) ~\phi^4$ theories. It is found that the disappearance
of tree-level threshold amplitudes of $2$ on-shell particles producing a
large number of particles occurs in $\phi^k$ theory only for $k=3$
and $k=4$. The one-loop correction to the threshold amplitude for a
highly virtual scalar particle decaying into $n$ particles in an
$O(N)$ model is derived.
\end{titlepage}

Some interesting properties of interactions involving large numbers or
particles in perturbation theory have become apparent in recent months.
Earlier in the year, the threshold amplitude for one highly virtual scalar
particle decaying into $n$ scalar particles was solved explicitly at
the tree-level using recursion relations$^{\cite{voloshin1,akp1}}$. In
the approach of Argyres, Kleiss, and Papadopoulos, a generating
functional was used to solve for the threshold amplitudes.

An approach using the reduction formula technique was suggested by
Brown$^{\cite{brown}}$ which shed some light on the physical
interpretation of the generating functonal. Brown's approach elegantly
reproduced previous results while showing the relation between the
generating functional and the solution of the classical field equations.

Brown's method was extended to the one-loop level for $\lambda\phi^4$
theories with broken and unbroken
reflection symmetry$^{\cite{voloshin3,smith}}$. In solving for these
one-loop corrections, explicit expressions for two-point Green functions
in the presence of a classical background field at the kinematical
threshold were derived. These Green functions enabled one to see the
surprising property that the amplitude for 2 incoming on-shell scalar
particles going to $n$ on shell scalar particles at rest vanishes for
some $n>N$. This $N$ was found to be 4 for the case of unbroken symmetry
and 2 for the case of broken symmetry. This cancelation of
diagrams with $n>N$ hints at some yet to be discovered symmetry. Voloshin
extended this discussion to the linear sigma model as well as the
production of Higgs bosons in the standard model by two incoming
fermions or gauge bosons$^{\cite{voloshin4}}$.

In this paper, we consider multiparticle production in $\phi^k$ and
$O(N) ~\phi^4$ theories. It is found that in $\phi^k$ theories, the
two-point Green function contains two double poles for $k=4$ and three
double poles for $k=3$ which correspond to the on-mass-shell scattering
amplitude of the process $2 \to n$. The vanishing of higher poles does
not occur for $k>4$.
\\

The first theory to be discussed is $\phi^k$. The Lagrangian for this
theory is

\beq
{\cal L} = {1\over{2}}(\partial\phi)^2 - {1\over{2}}m^2\phi^2 -
{\lambda\over{k}}\phi^k.
\label{phiklangrange}
\eeq
As discussed by Brown$^{\cite{brown}}$, we can use the reduction formula
method to write the threshold amplitude of 1 particle decaying into $n$
particles as

\beq
\langle n|\phi|0\rangle = \left [\prod_{a=1}^n \lim_{p_a^2 \to m^2}\int
d^4 x_a e^{i p_a x_a} \left (m^2-p_a^2\right )
{\delta\over{\delta\rho(x_a)}}\right ] \langle
0+|\phi(x)|0-\rangle^{\rho}|_{\rho=0}.
\label{reducform}
\eeq
where $\langle 0+|\phi(x)|0-\rangle^{\rho}\equiv \phi(x)$ is the
response of the field with a source term, $\rho\phi$ added to the
Lagrangian. At the kinematical threshold, all spatial $p_a$'s are equal
to zero, and $\rho(x)$ and $\phi(x)$ become
functions of $t$ only. At the tree level, the response can be replaced by
the solution of the classical field equations, $\phi_{cl}$. Eq.
($\ref{reducform}$) can be simplified by rewriting  it in terms of the
solution of the classical equations of motion in the limit as $\lambda
\to 0$,

\beq
z(t) \equiv {\rho(0) e^{i\omega t}\over{m^2-\omega^2-i\epsilon}}.
\label{zdef}
\eeq
This leaves us with the compact form for Eq. ($\ref{reducform}$),

\beq
\langle n|\phi|0\rangle = \left ( {\partial\over{\partial z}}\right )^n
\phi(z)|_{z=0}.
 \label{genfun}
\eeq

We can find $\phi_{cl}$ by solving the classical equations of motion,

\beq
{\partial^2 \phi_{cl}\over{\partial t}^2} + m^2 \phi_{cl}
+\lambda\phi_{cl}^{k-1} = 0.
 \label{classfield}
\eeq
It can be verified that the appropriate solution is

\beq
\phi_{cl} = {z(t)\over{[1-({\lambda/2km^2})z(t)^{k-2}]^{2/(k-2)}}}.
\label{classphik}
\eeq
This can be expanded to reproduce previous
results$^{\cite{voloshin1,akp1,brown}}$.

To find the first loop correction, $\phi(x)$ is replaced with its mean
value. We expand around the classical solution by writing $\phi(x) =
\phi_{cl}(x) + \phi_{1}(x)$ where $\phi_{1}(x)$ is the mean value of the
quantum part of the field. When this expansion is carried out
and the classical equations of motion are used, it is found
that $\phi_{1}$ must satisfy

\beq
{\partial^2 \phi_{1}(t)\over{\partial t}^2} + m^2 \phi_{1}(t)
+ (k-1)\lambda\phi_{cl}^{k-2}\phi_{1}(t)+
{1\over{2}}(k-1)(k-2)\lambda\phi_{cl}^{k-3}
\langle\phi_q(x)\phi_q(x)\rangle = 0,
\label{quantfield}
\eeq
where $\langle\phi_q(x)\phi_q(x)\rangle$ is the $\bf x \to x'$ limit
of the two-point Green function, $G(x,x')  \equiv\langle
T\phi_q(x)\phi_q(x)\rangle$, in the background of the classical
field. This Green function is the inverse of
the second variation of the Lagrangian evaluated in the classical
background field,

\beq
\left [\partial^{2}_{x} + m^2 + (k-1)\lambda\phi_{cl}(t)^{k-2} \right ]
G(x,x') = -i\delta(t-t')\delta^3({\bf x-x'}).
\label{greenform}
\eeq

The poles of the operator in eq. ($\ref{greenform}$) may be kept off
the real axis by analytic continuation into Euclidean space. It is
convienent to work with the quantity $u(\tau) \equiv e^{m\tau}$ defined
as, \beq
u(\tau) = z \left [{-\lambda\over{2km^2}}\right ]^{1/(k-2)}.
\label{udef}
\eeq
With this substitution, the operator in eq. ($\ref{greenform}$) takes
on the simple form,
\beq
{\partial^2\over{\partial\tau^2}}-\omega^2+{k(k-1) m^2\over{2
\cosh^{2}[{(k-2)m\tau/2}]}},
\label{greenop}
\eeq
where $\omega \equiv \sqrt{{\bf k}^{2}+m^2}$ and $\bf k$ is the
conserved spatial momenta. This operator is well known
and the solution has been discussed in Quantum
Mechanics$^{\cite{landau}}$ as well as previous works on this
subject$^{\cite{voloshin3,smith}}$. The homogeneous equation with this
operator has two solutions,

\beq
f_{\pm} = u^{\mp\omega/m} F(-s,s+1;1\pm 2\omega/(n-2);
(1-\tanh{[(k-2)m\tau/2]})/2),
\label{homogsolution}
\eeq
where $F(a,b;c;z)$ is the hypergeometric function and $s(s+1) \equiv
2k(k-1)/(k-2)^2$. The solutions reduce to a simple polynomial series
with a finite number of terms for all integer values of $s$. The finite
integers values of $k$ that yield integer values of $s$ are $k=3$ and
$k=4$. The case of $k=4$ has been discussed in previous
works$^{\cite{voloshin3,smith}}$, so we will continue this discussion
for the case $k=3$, i.e. $\phi^3$ theory.

In $\phi^3$ theory, the homogenous solutions in eq.
($\ref{homogsolution}$) are found to be simple polynomials. When the
Green function is constructed, it is found to have poles at $\omega
= \pm 1/2$, $\omega = \pm 1$, and $\omega = \pm 3/2$. There are no
poles at any higher integers. Thus, the tree-level threshold amplitude
for two on-shell scalar particles decaying into $n$ scalars must vanish
for all real processes except for $2 \to 3$.

When the Green function is summed over all values of ${\bf k}$, one
arrives at an expression for the two point Green function,
\beq
\langle\phi_q(x)\phi_q(x)\rangle = {I_{1}\over{2}} + {3
m^{2} u\over{(1+u)^2}} \left ( I_{3} + {1\over{2\pi^2}} -
{\sqrt{3}\over{12\pi}} \right )
+{5\sqrt{3}m^{2}u^{2}\over{2\pi(1+u)^{4}}} - {G
m^{2}u^{3}\over{(1+u)^6}},
\label{twoptexp}
\eeq
where
\beq
G \equiv {5\over{4\pi^2}} \left [ 3 \sqrt{5}
\ln{3+\sqrt{5}\over{3-\sqrt{5}}}-6 i \sqrt{5}\pi + 5\sqrt{3}\pi
\right ],
\label{Gdef}
\eeq
and $I_1$ and $I_3$ are divergent integrals. All but the last term of
eq. ($\ref{twoptexp}$) may be incorporated into the renormalized
coupling constants. The appropriate choice of renormalized values to
simplify eq. ($\ref{twoptexp}$) is
\begin{eqnarray}
\bar{m}^{2} = m^2 -{\lambda^{2}\over{2m^{2}}} \left
[ {I_{3}\over{2}} +{1\over{2\pi^2}} - {\sqrt{3}\over{12\pi}} \right ]
\nonumber \\
\bar{\lambda} = \lambda + {\lambda^{3}5\sqrt{3}\over{72m^{2}\pi}}
\label{renormalizedValues}
\end{eqnarray}

  Using eq. ($\ref{twoptexp}$), one can easily solve
eq. ($\ref{quantfield}$) to find
\beq
\phi(z(t)) = {z(t)\over{[1-(\lambda /6m^{2})z(t)]^2}} \left [ 1 -
{\lambda\over{8}} \left ( {\lambda\over{6m^2}} \right )^{3} {G
z(t)^2 \over{[1+(\lambda /6m^2)z(t)]^2}} \right ]
\label{phi3solution}
\eeq
When eq. ($\ref{phi3solution}$) is expanded, one finds the threshold
amplitude for $1 \to n$ processes in $\phi^3$ theory,
\beq
\langle n|\phi(0)|0\rangle = n \left ( {\bar{\lambda}\over{6\bar{m}^2}}
\right )^{n-1} n! \left [ 1 - {\lambda^{2}\over{288 m^2}} G (n-1)(n-2) \right
].
\label{phi3amp}
\eeq
\\

We now proceed to considering an $O(N) \phi^4$ theory. For an $N$
component theory invariant under $O(N)$ transformations, the Lagrangian
is

\beq
{\cal L} = {1\over{2}} (\partial_{\mu}\phi^a)(\partial^{\mu}\phi^a) -
{m^{2}\over{2}}\phi^{a}\phi^{a} -
{\lambda\over{4}}(\phi^{a}\phi^{a})^{2},
 \label{onlagrange}
\eeq
where $a$ is summed from $1$ to $N$. In this
theory, we can can define the solution to the free equations of motions
as $z^a(t)$ as $n^{a}z(t)$ where z(t) is defined in eq. ($\ref{zdef}$).
The classical solution can easily be found to be$^{\cite{brown}}$,
\beq
\phi^{a}_{cl}(z^{a}(t)) = {z^a(t)\over{1-(\lambda/8m^2)z(t)^2}}
\label{onclass}
\eeq

One can expand around the classical solution as was done for $\phi^k$
theory. The two point Green function,
$\langle\phi_{q}^{a}(x)\phi_{q}^{b}(x)\rangle$, is now a rank two
tensor. As before, $G^{ab}(x,x')$ is the inverse of the second
variation of the Lagrangian,
\beq
\left [ (\partial_{x}^{2}+m^2+\lambda\phi_{cl}^2)\delta^{ab}
+2\lambda\phi^{a}\phi^{b}\right ] G^{bc}(x,x') =
-i\delta^{ac}\delta^{4}(x-x')
\label{ongreeneqn}
\eeq
This Green tensor can be convienently written as the sum of two tensors,
\beq
G^{ab}(x,x') = n^{a}n^{b}G_{1}(x,x')+\left (
\delta^{ab}-n^{a}n^{b}\right )G_{2}(x,x')
\label{ongreenform}
\eeq

If eq. ($\ref{ongreeneqn}$) is contracted with the tensor
$n^{a}n^{c}$, one finds that $G_{1}(x,x')$ is the Green function for
 ordinary $\lambda\phi^4$ theory, i.e. the case
$N=1$ which has been discussed by Voloshin$^{\cite{voloshin3}}$.
This is the familiar solution of

\beq
\left [{\partial^{2}\over{\partial\tau^{2}}}
-\omega^2+{6\over{\cosh^2{m\tau}}} \right ] G_{1}(x,x') =
-i\delta^{4}(x-x').
\label{G1eqn}
\eeq
The solution of the homogeneous equation with the operator in eq.
($\ref{G1eqn}$) is eq. ($\ref{homogsolution}$) evaluated with $s=2$ and
$k=4$. In the equal time limit, $G_1$ becomes,
\beq
G_{1}(\tau,\tau) =
{1\over{2\omega}} + {6u^{2}\over{(1+u^{2})^{2}\omega^3}} +
{6 m^{4}u^{4}
[-4m^{2}(1+u^{2})^{2}+\omega^{2}(1+14u^{2}+u^{4})]\over{\omega^3
(\omega^{2}-4m^{2})(\omega^{2}-m^{2})(1+u^2)^4}}
 \label{green1}
\eeq
The solution for
$G_{1}(x,x')$ has poles only at $\omega =\pm 2m$ and
$\omega =\pm 4m$.

The other contribution to the Green tensor, $G_{2}$ can be found by
taking the trace of eq. ($\ref{ongreeneqn}$). With the help of eq.
($\ref{G1eqn}$) it is found that
\beq
\left [{\partial^{2}\over{\partial\tau^{2}}}
-\omega^2+{2\over{\cosh^2{m\tau}}} \right ] G_{2}(x,x') =
-iN\delta^{4}(x-x').
\label{G2eqn}
\eeq
This is the familiar nonreflective potential found when searching for
many of these Green functions, and the homogeneous equation has the
solution given by eq. ($\ref{homogsolution}$) with $s=1$ and $n=4$.
When $G_{2}$ is constructed, it is easy to show that it has poles at
$\omega =\pm m$. The equal time Green function is
\beq
G_{2}(\tau,\tau) = {1\over{2\omega}} + {2m^{2}u^{2}\over{(1+u^{2})^{2}
\omega (\omega^{2}-m^{2})}}
\label{G2}
\eeq

The Green functions needed to find the one-loop correction are found by
summing $G^{ab}(x,x)$ for all values of ${\bf k}$. One can easily
integrate over $G_1$ and $G_2$ to find,
\beq
\langle\phi^{a}_{q}\phi^{b}_{q}\rangle = \left [{I_{1}\over{2}}
+{2m^{2}u^{2}\over{(1+u)^{2}}}\left
(I_{3}+{1\over{2\pi^2}} \right )  \right ]
\delta^{ab} + \left [ {4m^{2}u^{2}\over{(1+u^{2})^2}} \left
(I_{3}+{1\over{2\pi^2}} \right ) -{6
u^{4}m^{2}F\over{(1+u^{2})^4}} \right ] n^{a}n^{b},
\label{onfinalgreen}
\eeq
where
\beq
F = {\sqrt{3}\over{2\pi^2}}\left ( \ln{2+\sqrt{3}\over{2-\sqrt{3}}}-i\pi
\right ),
\label{F}
\eeq
and $I_{1}$ and $I_3$ are divergent integrals. The Green funcition,
$\langle\phi^{a}_{q}\phi^{b}_{q}\rangle$, can be simplified to a single
finite term with the renormalization procedure,
\begin{eqnarray}
\bar{m}^2 = m^2 + {\lambda I_{1}(N+2)\over{2}} \nonumber
\\
\bar{\lambda} = \lambda - {\lambda^{2}(N+8)\over{4}} \left [
I_{3} +{1\over{2\pi^2}} \right ]
\label{ONrenormalizedValues}
\end{eqnarray}

This solution may be used in the expanded equation of motion to find
the value of $\phi^{a}(x)$ at the one-loop level,
\beq
\phi^{a}(z^{a}(t)) = {z^{a}(t)\over{1-(\bar{\lambda}/8\bar{m}^2)z(t)^2}}
\left [ 1- {3\lambda^{3}F z(t)^{4}\over{256
m^{4}(1-(\lambda/8m^{2})z(t)^2)}} \right ].
\label{onfield}
\eeq

As in $\phi^4$ theory, the one-loop correction to the $1 \to n$
amplitude in $\phi^3$ theory is found to grow as $n^2$. In $\phi^3$
theory, the tree-level threshold amplitude for $2 \to n$ processes is
found to vanish for all $n>3$. The vanishing of $2 \to n$ processes does
not occur in $\phi^k$ theory with $k>4$.

In $N$ component theories with an $O(N)$ symmetry, the one-loop
correction to the amplitude of one particle producing a total of $n$
particles at rest grows as $n^2$. The amplitude does not acquire an
explicit $N$ dependence, except for in the renormalization of $m^2$
and $\lambda$.

 I would like to thank Mikhail Voloshin for many
enlightening discussions.  This work was supported, in part, by
Department of Energy grant DOE-DE-AC02-83ER40105.


\begin{thebibliography} {99}
\bibitem{voloshin1} M.B. Voloshin, Minnesota
preprint TPI-MINN-92/1-T, January 1992, Nucl. Phys. B, to be published.
\bibitem{akp1}
E.N. Argyres, R.H.P Kleiss, and C.G. Papadopoulos, CERN preprint
CERN-TH-6496.
\bibitem{brown} L.S. Brown,
University of Washington preprint UW/PT-92-16, September 1992, to be
published in Phys. Rev. D.
\bibitem{voloshin3} M.B. Voloshin, Minnesota preprint TPI-MINN-92/45-T,
September 1992, submitted to Phys. Rev. D.
\bibitem{smith} B.H. Smith, Minnesota preprint TPI-MINN-92/50-T,
September 1992, submitted to Phys. Rev. D.
\bibitem{voloshin4} M.B. Voloshin, Minnesota preprint TPI-MINN-92/56-T,
October 1992.
\bibitem{landau} L.D. Landau
and E.M. Lifshitz, Quantum Mechanics, Non-Relativistic Theory, Third
edition,\S 23, Pergamon Press.
\end{thebibliography}
\end{document}